\documentclass[9pt,journal,twoside,web]{ieeecolor2}
\usepackage{generic}
\usepackage{cite}
\usepackage{amsmath,amssymb,amsfonts,mathtools}
\usepackage[linesnumbered,ruled,vlined]{algorithm2e}
\usepackage{graphicx}
\usepackage{textcomp}
\usepackage{multirow,booktabs}
\usepackage{hyperref}
\def\BibTeX{{\rm B\kern-.05em{\sc i\kern-.025em b}\kern-.08em T\kern-.1667em\lower.7ex\hbox{E}\kern-.125emX}}
\begin{document}
\bstctlcite{IEEE:BSTcontrol}
\title{Joint Poisson Reconstruction for Prompt-Gamma-Enabled $^{68}$Ga/$^{18}$F Dual-Isotope Quantitative PET Imaging}
\author{
Yu Shi, \IEEEmembership{Member, IEEE}, Meng Niu, Sen Yang, Tengjiao Wang, Risheng Yang, Yanyun Liu, Wei Ding, Xiangyu Liu, Fei Kang, and Shouping Zhu, \IEEEmembership{Member, IEEE}
\thanks{This work was supported by the National Natural Science Foundation of China (Nos. 62501415, 62027901, 62471372, and 62071362), the Natural Science Foundation of the Jiangsu Higher Education Institutions of China (No. 24KJD310001), the Natural Science Basic Research Program of Shaanxi (No. 2026JC-JCQN-093), the Natural Science Foundation of Jiangsu Province of China (No. BK20260182), the Jiangsu Government Scholarship for Overseas Studies, and the Qinglan Project of Jiangsu Province of China in 2024. Corresponding authors are Fei Kang (fmmukangf@fmmu.edu.cn) and Shouping Zhu (spzhu@xidian.edu.cn).}%
\thanks{Yu Shi and Wei Ding are with the School of Electrical Engineering and Automation, Suzhou University of Technology, Suzhou, Jiangsu 215500, China.}%
\thanks{Yu Shi, Sen Yang, Tengjiao Wang, Yanyun Liu, Xiangyu Liu, and Shouping Zhu are with the School of Life Science and Technology, Xidian University and Engineering Research Center of Molecular and Neuro Imaging, Ministry of Education, Xi'an, Shaanxi 710126, China.}%
\thanks{Meng Niu, Risheng Yang, and Fei Kang are with the Department of Nuclear Medicine, Xijing Hospital, Fourth Military Medical University, Xi'an, Shaanxi 710032, China.}}

\maketitle

\begin{abstract}
\textcolor{cyan}{\textit{Objective:}}
To enable simultaneous quantitative $^{68}$Ga/$^{18}$F PET imaging despite low prompt-gamma triple-coincidence counts and random-triple contamination.
\textcolor{cyan}{\textit{Methods:}}
We developed an expectation-maximization reconstruction method (PNR-EM) that jointly estimates pure-emitter, non-pure-emitter, and random-triple components using a Poisson model of prompt double, prompt triple, and delayed triple measurements.
A consistent-delay strategy estimates random triples while preserving true-pair temporal correlations. Prompt doubles additionally constrain the non-pure component, and total variation regularization suppresses reconstruction noise. The method was evaluated with quantitative and Derenzo phantoms, a nine-well separation phantom, and tumor-bearing mice.
\textcolor{cyan}{\textit{Results:}}
Phantom quantitation showed strong linearity ($r^2>0.99$) and a maximum relative deviation below 9.9\% relative to single-isotope references. The separated images largely preserved rod resolvability relative to their corresponding single-isotope references. The best PSNR, SSIM, and RMSE were 33.72~dB, 0.975, and 0.0206 for $^{18}$F, and 34.86~dB, 0.955, and 0.0181 for $^{68}$Ga in the 9-well phantom. The $^{68}$Ga RMSE was 41.4\% lower than that of the best image-domain separation baseline. Tumor-bearing mouse imaging demonstrated in vivo separation of $^{68}$Ga-DOTA-NOC and $^{18}$F-FDG distributions.
\textcolor{cyan}{\textit{Conclusion:}}
Joint estimation improved dual-isotope separation under the evaluated low-count conditions.
\textit{Significance:}
To our knowledge, this study provides the first experimental demonstration of simultaneous $^{68}$Ga/$^{18}$F PET separation using prompt-gamma coincidences, supporting simultaneous imaging of complementary metabolic and receptor-targeted tracer distributions.
\end{abstract}

\section{Introduction}

Positron Emission Tomography (PET) enables non-invasive quantitative imaging of biochemical and physiological processes in vivo, making it a vital component of the standard imaging technique for oncology diagnosis \cite{RN1}. PET imaging relies on radiotracers, of which $^{18}$F-FDG is the most widely used radiotracer for cancer staging and precise localization of metastatic lesions \cite{RN2}. However, single isotope $^{18}$F-FDG only provides information on glucose metabolism, leading to potential false positives in benign proliferative lesions (e.g., nonspecific lymph node uptake \cite{RN3}), and false negatives in certain malignant neoplasms with high uptake background (e.g., hepatocellular carcinoma \cite{RN4}). In recent years, $^{68}$Ga-labeled tracers have undergone rapid development and shown complementary effects to $^{18}$F radiopharmaceuticals \cite{RN5,RN6}, addressing the limitations of $^{18}$F-FDG in disease evaluation. For instance, the combined image information of $^{18}$F-FDG and $^{68}$Ga-labeled somatostatin analogs (e.g., DOTA-TATE, DOTA-NOC) is crucial for clinicians to map the entire extent of tumor cell differentiation, as well-differentiated neuroendocrine neoplasms (NENs) have a higher avidity for $^{68}$Ga-labeled somatostatin analogs, whereas poorly differentiated NENs show intense uptake of $^{18}$F-FDG \cite{RN8,RN9}. Therefore, the combination of $^{68}$Ga and $^{18}$F isotopes is expected to enhance diagnostic sensitivity and staging accuracy \cite{RN10}, and this strategy is incorporated into official diagnostic guidelines in some countries \cite{RN11, RN12}.

The physical basis of PET imaging involves the coincidence detection of a pair of 511 keV gamma photons from the same positron annihilation. However, annihilation photons generated by different positron-emitting radionuclides are intrinsically indistinguishable, making it challenging for static PET systems to discriminate between different nuclide signals. Therefore, conventional dual-tracer PET generally requires separate single-tracer acquisitions at different time points \cite{RN14}, which prolongs the imaging protocol, increases radiation exposure, and introduces uncertainties associated with patient repositioning, physiological changes, and tracer pharmacokinetics \cite{RN10,RN15}. 
From the perspective of nuclear imaging technology, simultaneous multi-isotope PET is not merely a clinical protocol issue, but also a challenging problem in radiation detection, coincidence acquisition, and quantitative reconstruction. Additional discriminative information beyond static double-coincidence measurement is required to discriminate different radionuclide signals within a single acquisition.

Recent methodological developments and potential clinical applications of simultaneous multitracer PET have been reviewed by Li et al. \cite{Li2026MultitracerPET}. Strategies for simultaneous dual-isotope PET imaging can be broadly categorized into two main approaches \cite{RN15}: dynamic-scanning-based methods \cite{RN19, RN20, RN21} and prompt-gamma-based methods \cite{RN22, RN23, RN24, RN25}. The former treats the dual-isotope data as mixed time-varying signals, then recovers single-tracer distributions by modeling differences in biodistribution kinetics and radioactive decay, using approaches such as multi-tracer compartment modeling \cite{RN19} and machine-learning-based signal separation \cite{RN10, RN21}. However, the multi-frame dynamic scanning protocol usually requires a long acquisition duration, complicated operation, and organ-dependent kinetic modeling \cite{RN25}, which may hinder its clinical translation. In contrast, prompt-gamma-based methods exploit the physical decay characteristics of non-pure positron emitters. Such radionuclides can emit two annihilation gammas ($\beta^+$) together with an additional prompt gamma ($\gamma$) almost simultaneously owing to nuclear de-excitation, thereby enabling $\beta^+\gamma$ triple-coincidence detection. This detector-level information provides a physical basis for distinguishing pure and non-pure emitters without time-consuming dynamic scanning. 
Recent studies have further demonstrated the feasibility of prompt-gamma-enabled multi-isotope PET. For example, multiplexed PET using $^{55}$Co/$^{18}$F has been investigated, in which triple coincidences from $^{55}$Co were used to distinguish it from the pure positron emitter $^{18}$F in a single imaging session \cite{Zou_2025}. In addition, maximum-likelihood-based line-of-response assignment has been proposed for triple coincidences by leveraging spatial, timing, and energy information, highlighting the importance of accurate event-level discrimination in prompt-gamma PET \cite{Zou_2026}. A recent reconstruction study further incorporated prompt-gamma attenuation to reduce the misclassification of $\beta^+\gamma$ events as conventional double-coincidence events \cite{Pfaehler_2025}. These advances indicate that prompt-gamma information is a promising physical basis for simultaneous multi-tracer PET imaging.

Considering that $^{18}$F is a pure emitter whereas $^{68}$Ga is a non-pure emitter, the prompt-gamma-based approach is theoretically well suited for simultaneous $^{68}$Ga/$^{18}$F imaging. However, in previous studies \cite{RN25, RN26, RN27}, $^{68}$Ga and its parent $^{68}$Ge were imaged as pure emitters because the prompt-gamma branch ratio of $^{68}$Ga is only 1.2\% \cite{RN28}, much lower than the $>10\%$ requirement for current dual-isotope separation methods \cite{RN15}. Researchers have pursued higher prompt-gamma detection efficiency to relax the constraint on candidate isotopes, and the efficiency of dual-isotope PET system has risen from 0.6\% \cite{RN27} to 7.0\% \cite{RN23} over the last decade. This progress is closely related to advances in PET detector design and front-end readout electronics, which are essential for acquiring reliable coincidence information in advanced nuclear imaging systems \cite{Goldschmidt2016TBME}. Statistical reconstruction and regularization have been extensively investigated for noisy and low-count PET data \cite{Zhang2018TBME,Xie2022TBME}. However, these approaches do not explicitly address isotope separation using prompt and delayed triple-coincidence measurements. Consequently, existing prompt-gamma-based separation approaches have largely relied on image subtraction \cite{RN25} or projection subtraction \cite{RN24}. However, the extremely low probability of prompt gamma emissions results in severe noise, bias and incorrect quantitation \cite{RN29} in the triple-coincidence image of $^{68}$Ga, thereby hindering simultaneous $^{68}$Ga/$^{18}$F dual-isotope quantitative PET imaging without the development of advanced separation algorithms.

In addition to the low-count nature of triple coincidences, random triple coincidences further degrade dual-tracer quantification. Although the separation of non-pure and pure emitters can be realized theoretically using triple coincidences, pure emitters can still generate random triple coincidences, making the prompt-gamma-based approach less effective or even invalid. The influence of random triples on dual-tracer imaging is more severe than that of random doubles on single-tracer PET, where randoms mainly reduce contrast and signal-to-noise ratio \cite{RN30}. Conventional random correction based on delayed window requires high statistical quality of data \cite{RN31} and uncorrelation among random photons \cite{RN32}. However, the correlated random triple (CRT), comprising a random $\gamma$ and a pair of true double coincidences from the same annihilation \cite{RN33}, constitutes 43.2\% of all triple coincidences in a $^{68}$Ga scan \cite{RN34}, highlighting the need to preserve the temporal correlation within the true-double component when estimating CRTs. Meanwhile, the low-count nature of triple coincidences weakens the statistical reliability of delayed-window estimation. Failure to account for this correlation, together with low delayed-triple counts, can compromise random-triple correction and subsequent isotope separation. To solve this problem, Moore et al. used conventional PET reconstruction results to post-correct CRT in the image domain by assuming a linear correlation between CRT and true double coincidence \cite{RN35}. However, this image-domain random correction relies on the high-quality triple coincidence images, which are difficult to obtain under the low-count conditions of triples in $^{68}$Ga scan. More recently, researchers extended delayed-window techniques to estimate random double and triple coincidences, accounting for two- and three-decay triple contributions and the effects of coincidence and veto windows
\cite{Niekamper2026RandomTriples}. These developments motivate further investigation of how random-triple observations can be incorporated into joint reconstruction for low-count $^{68}$Ga/$^{18}$F dual-isotope PET.

In summary, dual-tracer PET imaging techniques based on triple coincidence have achieved important progress in the past few years, potentially enabling simultaneous $^{68}$Ga/$^{18}$F dual-tracer imaging for applications such as histological diagnosis prediction \cite{RN5}. However, the extremely low prompt-gamma emission probability of $^{68}$Ga, together with random-triple contamination, substantially reduces quantitative accuracy in simultaneous $^{68}$Ga/$^{18}$F dual-tracer imaging. Therefore, a novel separation method is urgently needed to overcome the above challenges. This study aims to develop an algorithm for simultaneous dual-isotope quantitative PET imaging of $^{68}$Ga/$^{18}$F, focusing on low-probability and random-contaminated triple coincidences. Inspired by previous work \cite{RN29}, the proposed method integrates pure positron-emitter, non-pure-emitter, and random-triple components into a joint Poisson model. Additionally, total variation (TV) regularization is introduced to suppress noise. The constrained optimization problem is solved using the Green's one-step-late expectation maximization (EM) algorithm based on consistent delayed observation, thereby mitigating the influence of random contamination and low-count characteristics on the quantification of dual-isotope imaging. 
Phantom experiments evaluated quantitative reconstruction performance, and tumor-bearing mouse imaging assessed in vivo feasibility of simultaneous $^{68}$Ga/$^{18}$F dual-tracer separation.
The rest of this paper is organized as follows. In Section II, we present and derive the joint-Poisson-model-based reconstruction algorithm, termed the simultaneous Pure, Non-pure, and Random EM reconstruction algorithm (PNR-EM), followed by its implementation details. In Section III, the proposed algorithm is applied to a small-animal PET, where we evaluate algorithmic performance via physical phantom and in-vivo mouse experiments. Finally, the discussion and conclusion are provided in Sections IV and V, respectively.

\section{Methods}\label{sec:artwork}
\subsection{PNR-EM Reconstruction}

In PET imaging, the number of coincidence events generated by a radioactive source can be modeled as a Poisson process. Hence, dual-isotope PET imaging can be conceptually described by two source-specific Poisson models for pure and non-pure emitters, respectively \cite{RN29}, as shown in Eqs. \eqref{eq1} and \eqref{eq2}.
\begin{equation}\label{eq1}
	dp_A \sim \mathrm{Poisson}\left( {{f_A}} \right)
\end{equation}
\begin{equation}\label{eq2}
	 dp_B \sim \mathrm{Poisson}\left( {{f_B}} \right), \quad tp_B\propto dp_B
\end{equation} 
where $f_A$ denotes the spatial distribution of a pure emitter, and $f_B$ denotes the spatial distribution of a non-pure emitter in a dual-isotope acquisition. The terms $dp_A$ and $dp_B$ represent the source-specific prompt double-coincidence components from the pure and non-pure emitters, respectively. The term $tp_B$ denotes the ideal true triple-coincidence component generated by the non-pure emitter. Since $dp_B$ and $tp_B$ originate from the same non-pure emitter distribution, they are expected to be proportional under ideal conditions. However, for widely-used non-pure emitters such as $^{68}$Ga and $^{124}$I, random triple coincidences are non-negligible, thereby violating the above ideal assumption. As mentioned in Section I, the pre-correction using projection subtraction cannot effectively address the CRT problem due to limited triple-count statistics and the preserved intra-event correlation in CRTs. Here, we first employ a consistent delay (CD) strategy to estimate random triple coincidences.

\subsubsection{Consistent Delay for Random Triple Estimation}
\begin{figure}[!htb]
	\includegraphics
	[width=\hsize]
	{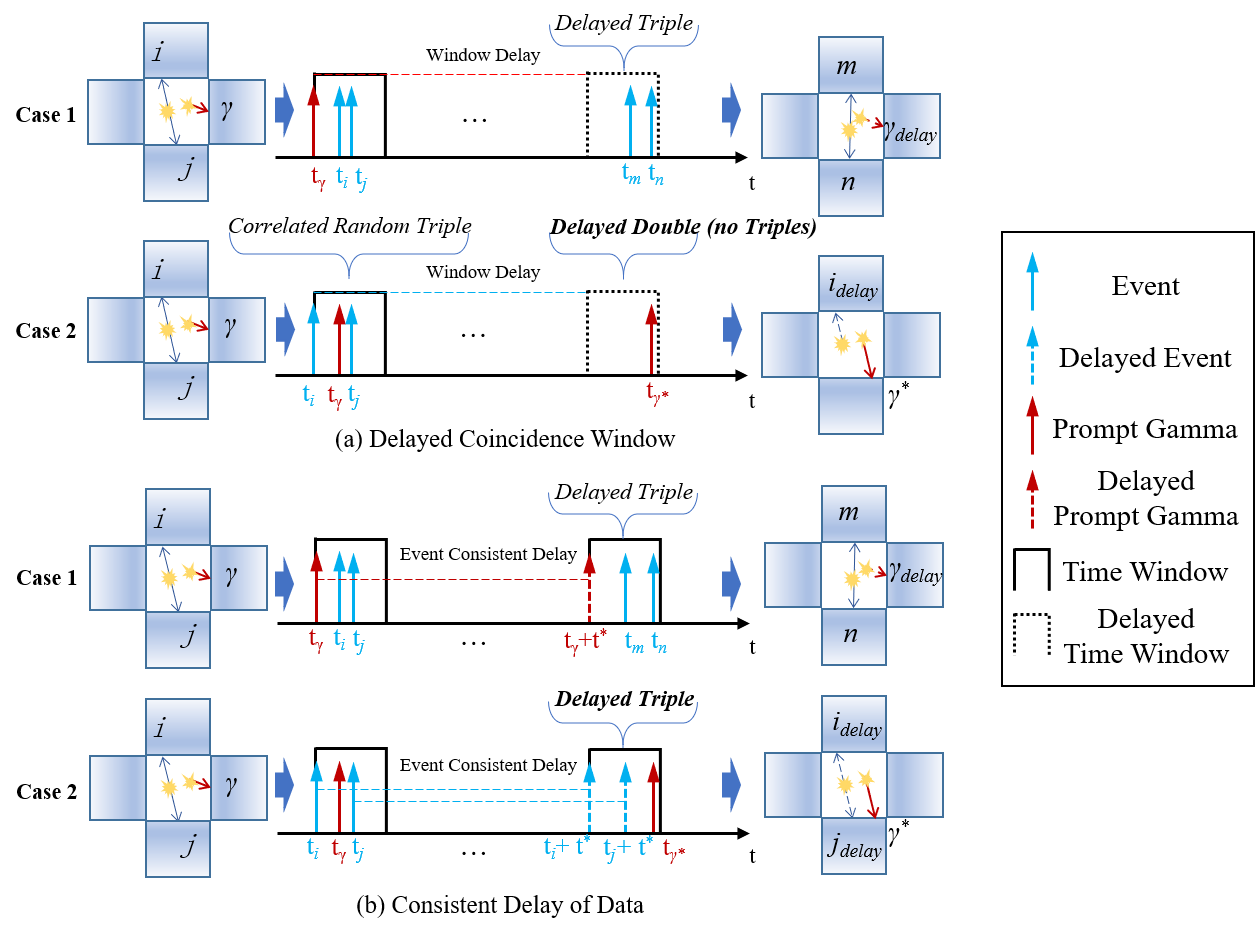}
	\caption{Schematic comparison of the channel-wise delayed-window implementation evaluated in this study and the consistent-delay strategy for estimating random triple coincidences.}
	\label{Fig1}
\end{figure}

To reliably estimate random triple coincidences, particularly CRTs, the single-photon event stream is delayed as a whole rather than by independently delaying individual detector channels. In this way, the relative timing between photons originating from the same decay is preserved. Therefore, the two annihilation photons constituting a true double coincidence remain temporally correlated in the delayed event stream, enabling random triples to be estimated by exploiting the statistical independence between different decay events. The detailed procedure of the proposed CD strategy is summarized in Algorithm 1. Briefly, a fixed delayed copy of the original single-photon event stream is first generated and then merged with the original stream. After time sorting and energy-window filtering, the same triple-coincidence sorting procedure used for prompt data is applied to extract delayed triples, as an estimate of random triple coincidences. 
As shown in \textbf{Case 1} of Fig. \ref{Fig1}, both the conventional delayed-window/channel (DW) method \cite{RN32, RN36} evaluated in this study and the proposed CD strategy can estimate uncorrelated random events in double- or triple-coincidence measurements. For CRTs, however, preserving the temporal correlation within the true-double component is essential. In the channel-wise DW implementation illustrated in \textbf{Case 2} of Fig. \ref{Fig1}(a), different delays applied to the two annihilation-photon channels disrupt this temporal relationship, leading to underestimation of the corresponding CRT contribution. In contrast, the CD strategy in Fig. \ref{Fig1}(b) shifts the event stream as a whole and preserves the relative timing of the true-double component, thereby enabling more appropriate estimation of CRTs.

\begin{algorithm}[!htb]
    \small
	\SetAlgoLined
	\KwIn{\\
		\hspace{10pt}Prompt single photon data $s_p(n)$, where $n = 1, 2,\ldots, N$,\\
		\hspace{10pt}Consistent delayed time $t^*$,\\
		\hspace{10pt}Coincidence time window $TW$,\\
		\hspace{10pt}Energy window for annihilation gammas $E_d$,\\
		\hspace{10pt}Energy window for prompt gammas $E_u$}
	\KwOut{Delayed triple counts $triple_d$ in each LOR, which are used as an estimate of random triples}
	Apply a consistent delay time $t^*$ to the single-photon data to generate the delayed event sequence $s_d(n)$, where $n=1,2,\ldots,N$.\;
	Merge $s_p(n)$ and $s_d(n)$, and sort the combined events according to their arrival times to obtain a time-ordered sequence $s_{all}(n)$, where $n=1,2,\ldots,2N$.\;
	Filter annihilation-gamma and prompt-$\gamma$ candidates using the energy windows $E_d$ and $E_u$, respectively, to generate the filtered sequence $s_{filter}(m)$, where $m=1,2,\ldots,M$.\;
	\For{$m=1$ \KwTo $M-2$}{
		\If{three sequential gammas fall within $TW$\\
        \& exactly two events satisfy the annihilation-gamma energy window $E_d$\\
		\& exactly one gamma satisfies the prompt-gamma energy window $E_u$\\
	\& the coincidence contains at least one prompt event and one delayed event}{Treat the coincidence as a delayed triple and record the LOR defined by the two annihilation gammas}
	}
	\caption{Consistent Delay for Estimating Random Triples}
\end{algorithm}

\subsubsection{Simultaneous Pure, Non-pure, and Random EM Reconstruction}
Using the CD strategy mentioned above, delayed triple coincidences are extracted before reconstruction and used as an estimate of random triple contamination. Under the event-selection rules used in this study, the raw delayed triple counts have the same expected counts as the prompt random triples:
$\mathbb{E}[td\mid f_r]=\mathbb{E}[tp_r\mid f_r]$.
The terms $td$ and $tp_r$ represent delayed triple coincidences and prompt random triple coincidences, respectively. Both components are modeled as Poisson counts with matching expected projection intensities, parameterized by the effective image-domain distribution $f_r$ in the complete-data formulation below.
The distribution $f_r$ is not interpreted as a physical radioactive distribution, but as an effective image-domain representation of random triple contamination after LOR binning. For brevity, the term "random emitter" is used throughout this paper to denote this effective image-domain component. The measured prompt triple coincidences consist of two components: the true triples $tp_B$ generated by the non-pure emitter and the prompt random triples $tp_r$. For computational tractability, the complete-data count contributions are modeled as conditionally independent Poisson variables given the image parameters $f_A$, $f_B$, and $f_r$. Under this approximation, the joint complete-data likelihood $L(f_A,f_B,f_r)$ is formulated as
\begin{equation}
\label{eq4}
\begin{aligned}
L(f_A,f_B,&f_r)
= \\\prod_{i,j} \Bigg(
& e^{-a_{ij}\omega_A f_A}
  \frac{\left(a_{ij}\omega_A f_A\right)^{dp_A}}{dp_A!} \\
& \times e^{-a_{ij}(1-\beta(j))\omega_B f_B}
  \frac{\left(a_{ij}(1-\beta(j))\omega_B f_B\right)^{dp_B}}{dp_B!} \\
& \times e^{-a_{ij}\beta(j)f_B}
  \frac{\left(a_{ij}\beta(j)f_B\right)^{tp_B}}{tp_B!} \\
& \times e^{-a_{ij}f_r}
  \frac{\left(a_{ij}f_r\right)^{tp_r}}{tp_r!} \\
& \times e^{-a_{ij}f_r}
  \frac{\left(a_{ij}f_r\right)^{td}}{td!}
\Bigg).
\end{aligned}
\end{equation}
where $i$ represents the bin index in the projection domain, and $j$ denotes the voxel index in the image domain. The term $a_{ij}$ corresponds to the system response matrix (SRM), while $\beta$ is the spatially variant normalization factor for triple coincidences. In this work, $\beta$ is considered as the product of the prompt-gamma emission probability and the prompt-gamma detection efficiency, which is estimated from $^{22}$Na experiments \cite{RN23}. The coefficients $\omega_A$ and $\omega_B$ denote the correction coefficients compensating for the true double coincidence that constitutes the CRT. In practice, since the number of CRTs is much smaller than that of true double coincidences, $(\omega_A,\omega_B)\approx1$. By omitting $\omega_A$, $\omega_B$, and constant terms, the approximate joint log-likelihood function is shown in Eq. (\ref{eq5}).
\begin{equation}
\label{eq5}
\begin{aligned}
\ln L(f_A,f_B&,f_r)= \\\sum\nolimits_{i,j} \Big(
 &-a_{ij}f_A(j)
 + dp_A(i,j)\ln\!\big(a_{ij}f_A(j)\big) \\
&
 -a_{ij}f_B(j)
 + dp_B(i,j)\ln\!\big(a_{ij}(1-\beta(j))f_B(j)\big) \\
&
 -a_{ij}f_r(j)
 + tp_r(i,j)\ln\!\big(a_{ij}f_r(j)\big) \\
&
 -a_{ij}f_r(j)
 + td(i,j)\ln\!\big(a_{ij}f_r(j)\big)\\
&
 + tp_B(i,j)\ln\!\big(a_{ij}\beta(j)f_B(j)\big)
\Big).
\end{aligned}
\end{equation}

Since the complete data ( $dp_A(i,j)$, $dp_B(i,j)$, $tp_B(i,j)$, $tp_r(i,j)$, and $td(i,j)$ ) are unobserved, their conditional expectations are estimated using the current images $f_A^k$, $f_B^k$, and $f_r^k$:
\begin{equation}\label{eq6}
\left\{ \begin{array}{l}
	dp_A({i,j}) = {double_p(i)}\frac{{a_{ij}}f_A^k(j)}{{\sum\limits_{j'} {{a_{ij'}}\left( {f_A^k(j') + ({1 - \beta(j')})f_B^k(j')} \right)} }}\\
	dp_B\left( {i,j} \right) = {double_p(i)}\frac{a_{ij}( 1-\beta(j))f_B^k(j)}{{\sum\limits_{j'} {a_{ij'}\left( {f_A^k(j') +( {1 - \beta(j')})f_B^k(j')} \right)} }}\\
	tp_B({i,j}) = triple_p(i)\frac{a_{ij}\beta(j)f_B^k(j)}{{\sum\limits_{j'} {a_{ij'}\left( {\beta(j')f_B^k(j') + f_r^k(j')} \right)} }}\\
	t{p_r}({i,j}) = triple_p(i)\frac{a_{ij}f_r^k(j)}{{\sum\limits_{j'} {a_{ij'}\left( {\beta(j')f_B^k(j') + f_r^k(j')} \right)} }}\\
	td({i,j}) = triple_d(i)\frac{a_{ij}f_r^k(j)}{{\sum\limits_{j'} {a_{ij'}f_r^k(j')} }}
\end{array} \right.
\end{equation}
where $double_p$, $triple_p$, and $triple_d$ denote the measured prompt double coincidences, prompt triple coincidences, and delayed triple coincidences, respectively. 

For notational simplicity, the complete-data symbols in the E-step and subsequent derivative expressions denote their conditional expectations at iteration $k$. In the M-step, $\ln L$ denotes the expected complete-data log-likelihood, with these expectations held fixed during differentiation with respect to $f_A$, $f_B$, and $f_r$. Let $s_j=\sum_i a_{ij}$ denote the sensitivity image. Setting the partial derivatives of the expected complete-data log-likelihood to zero yields the stationarity conditions in Eq. \ref{eq7}, from which the updates for the pure, non-pure, and random components are obtained.
\begin{equation}\label{eq7}
\left\{ \begin{aligned}
\frac{\partial \ln L}{\partial f_A(j)}
&=-s_j+\frac{1}{f_A(j)}\sum_i dp_A(i,j)=0,\\
\frac{\partial \ln L}{\partial f_B(j)}
&=-s_j+\frac{1}{f_B(j)}\sum_i\left(dp_B(i,j)+tp_B(i,j)\right)=0,\\
\frac{\partial \ln L}{\partial f_r(j)}
&=-2s_j+\frac{1}{f_r(j)}\sum_i\left(tp_r(i,j)+td(i,j)\right)=0.
\end{aligned} \right.
\end{equation}

Further, to suppress severe noise caused by low-count conditions, TV regularization is also introduced. As presented in Eq. (\ref{eq8}), the final iterative formulas with TV denoising are solved using the Green's one-step-late algorithm \cite{Green1990ModifiedEM}, where $U_A^k(j)$, $U_B^k(j)$, and $U_r^k(j)$ represent the partial derivatives of TV penalty for pure, non-pure and random images, respectively. $\lambda$ denotes the weight factor of regularization.
  \begingroup\small\begin{equation}\label{eq8}
	\left\{ \begin{aligned}
		f_A^{k + 1}&(j) =\frac{{f_A^k\left( j \right)}}{{s_j  + \lambda U_A^k\left( j \right)}}\cdot\\&\sum\limits_i {\frac{{{a_{ij}}double_p\left( i \right)}}{{\sum\limits_{j'} {{a_{ij'}}\left( {f_A^k(j') + ({1 - \beta (j')})f_B^k(j')} \right)} }}} \\
		f_B^{k + 1}&(j)=\frac{{f_B^k\left( j \right)}}{{s_j  + \lambda U_B^k\left( j \right)}}\cdot\\&\left( \begin{array}{l}
			\left( {1 - \beta \left( j \right)} \right)\sum\limits_i {\frac{{{a_{ij}}double_p\left( i \right)}}{{\sum\limits_{j'} {{a_{ij'}}\left( {f_A^k\left( j' \right) + \left( {1 - \beta \left( j' \right)} \right)f_B^k\left( j' \right)} \right)} }}} \\
			+ \beta \left( j \right)\sum\limits_i {\frac{{{a_{ij}}triple_p\left( i \right)}}{{\sum\limits_{j'} {{a_{ij'}}\left( {\beta \left( j' \right)f_B^k\left( j' \right) + f_r^k\left( j' \right)} \right)} }}} 
		\end{array} \right)\\
		f_r^{k + 1}&(j) =\frac{{f_r^k\left( j \right)}}{{2s_j  + \lambda U_r^k\left( j \right)}}\cdot\\&\sum\limits_i {\left( {\frac{{{a_{ij}}triple_p\left( i \right)}}{{\sum\limits_{j'} {{a_{ij'}}\left( {\beta \left( j' \right)f_B^k\left( j' \right) + f_r^k\left( j' \right)} \right)} }} + \frac{{{a_{ij}}triple_d\left( i \right)}}{{\sum\limits_{j'} {{a_{ij'}}f_r^k\left( j' \right)} }}} \right)} 
	\end{aligned} \right.
\end{equation}\endgroup
The above approach for $^{68}$Ga/$^{18}$F dual-isotope imaging is termed the simultaneous Pure, Non-pure, and Random EM reconstruction algorithm (PNR-EM).

\subsubsection{Algorithm Implementation}
The implementation of the proposed PNR-EM is outlined in Algorithm 2. The preprocessing step consists of two parts: (1) coincidence extraction with an energy window of 350--650 keV for annihilation gammas and 800--2000 keV for prompt gammas, and (2) correction of double coincidences, including random correction, normalization and gap-filling \cite{RN37} in turn. In the first step, the normalization factor for triple coincidences $\beta$ is set to one, and the algorithm degenerates into simultaneous Non-pure and Random EM reconstruction (NR-EM). After several iterations, the preliminary estimations of the non-pure and random components are obtained rapidly. A simple image-domain separation follows to obtain an initial estimate of the pure emitter. In the second step, $\beta$ is set according to the $^{22}$Na experiment constructed in advance, and the distributions of pure, non-pure and random emitters are obtained through an iterative loop of Eq. (\ref{eq8}) based on the initial values achieved in Step 1. The proposed reconstruction algorithm was implemented in C++ and CUDA on an NVIDIA RTX 3080 GPU. In this work, the number of NR-EM pre-reconstruction iterations was empirically set to 20 to accelerate the algorithm convergence.

\begin{algorithm*}[!htb]
    \small
	\SetAlgoLined
	\KwIn{\\
		\hspace{10pt}Corrected prompt double-coincidence measurement $double_p$, \\
        \hspace{10pt}Prompt triple-coincidence measurement $triple_p$, \\
        \hspace{10pt}Delayed triple-coincidence measurement $triple_d$ obtained using Consistent Delay,\\
		\hspace{10pt}Normalization factor for triple coincidences $\beta$,\\
		\hspace{10pt}Maximum number of NR-EM pre-reconstruction iterations $L$,\\
		\hspace{10pt}Maximum number of PNR-EM iterations $K$,\\
		\hspace{10pt}Weight factor of regularization $\lambda$}
	\KwOut{$f_A^K$, $f_B^K$, $f_r^K$}
    Compute the sensitivity image $s_j=\sum_i a_{ij}$ for all voxels $j$\;
    
    \textbf{NR-EM pre-reconstruction:}\\ Set $\beta=\mathbf{1}$, $\tilde f_B^0=\mathbf{1}$, and $\tilde f_r^0=\mathbf{1}$\;
	\For{$l=0$ \KwTo $L-1$}{
        Compute the forward projections $\hat t_i^l=\sum_{j'}a_{ij'}\left[\tilde f_B^l(j')+\tilde f_r^l(j')\right], \quad \hat r_i^l=\sum_{j'}a_{ij'}\tilde f_r^l(j') $\;
        
        Update the non-pure component: $\tilde f_B^{l+1}(j) = \frac{\tilde f_B^l(j)} {s_j+\lambda U_B^l(j)} \sum_i \frac{a_{ij}triple_p(i)} {\hat t_i^l} $\;
        
		Update the random component: $\tilde f_r^{l+1}(j) = \frac{\tilde f_r^l(j)} {2s_j+\lambda U_r^l(j)} \sum_i \left[ \frac{a_{ij}triple_p(i)} {\hat t_i^l} + \frac{a_{ij}triple_d(i)} {\hat r_i^l} \right]$ \;}
		Set $\beta$ according to the $^{22}$Na calibration experiment constructed in advance\;
		Initialize estimations on non-pure and random components by $f_B^0(j)=\frac{\tilde f_B^L(j)}{\beta(j)}, \quad f_r^0(j)=\tilde f_r^L(j)$\;
		Obtain initial estimations on pure emitter $f_A^0$ by image-domain separation for dual-isotope imaging;\\
		\textbf{Joint PNR-EM reconstruction:}\\ \For{$k=0$ \KwTo $K-1$}{
            Compute the forward projections $\hat d_i^k = \sum_{j'}a_{ij'} \left[ f_A^k(j')+(1-\beta(j'))f_B^k(j') \right], \hat t_i^k = \sum_{j'}a_{ij'} \left[ \beta(j')f_B^k(j')+f_r^k(j') \right], \quad \hat r_i^k = \sum_{j'}a_{ij'}f_r^k(j') $\;

            Update the pure emitter: $ f_A^{k+1}(j) = \frac{f_A^k(j)} {s_j+\lambda U_A^k(j)} \sum_i \frac{a_{ij}double_p(i)} {\hat d_i^k} $\;
            
			Update the non-pure emitter: $ f_B^{k+1}(j) = \frac{f_B^k(j)} {s_j+\lambda U_B^k(j)} \left[ (1-\beta(j)) \sum_i \frac{a_{ij}double_p(i)} {\hat d_i^k} + \beta(j) \sum_i \frac{a_{ij}triple_p(i)} {\hat t_i^k} \right] $\;
            
			Update the random component: $ f_r^{k+1}(j) = \frac{f_r^k(j)} {2s_j+\lambda U_r^k(j)} \sum_i \left[ \frac{a_{ij}triple_p(i)} {\hat t_i^k} + \frac{a_{ij}triple_d(i)} {\hat r_i^k} \right]$ \;
			}
	\caption{Pseudo code of simultaneous Pure, Non-pure, and Random EM reconstruction}
\end{algorithm*}

\subsection{Experimental Setup}
Four experiments were designed to evaluate the feasibility and quantification performance of PNR-EM algorithm. These experiments included: (1) a $^{68}$Ga/$^{18}$F quantitation assessment on 96-well plate phantoms with fixed well dimensions, (2) an evaluation of the effect of the proposed algorithm on image resolution with fixed activity, (3) a physical phantom study for comparing different dual-isotope separation methods, and (4) an in-vivo imaging of tumor-bearing mice. All experiments were conducted using the custom-built small-animal PET system\cite{RN38}. The SRM $a_{ij}$ was constructed using the multi-ray method \cite{RN39}. The image dimension was set to 216 $\times$ 216 $\times$ 292, with an isotropic voxel of 0.5 $\times$ 0.5 $\times$ 0.5 mm$^3$, according to 2.00 $\times$ 2.00 $\times$ 13.00 mm$^3$ LYSO crystal pitch. To ensure a fair comparison, the same TV regularization weight ($\lambda=3\times10^{-5}$) was used for all applicable methods unless otherwise stated.

\subsubsection{Quantitation Assessment on a 96-Well Plate}
\begin{figure}[!tbp]
	\includegraphics
	[width=1.0\hsize]
	{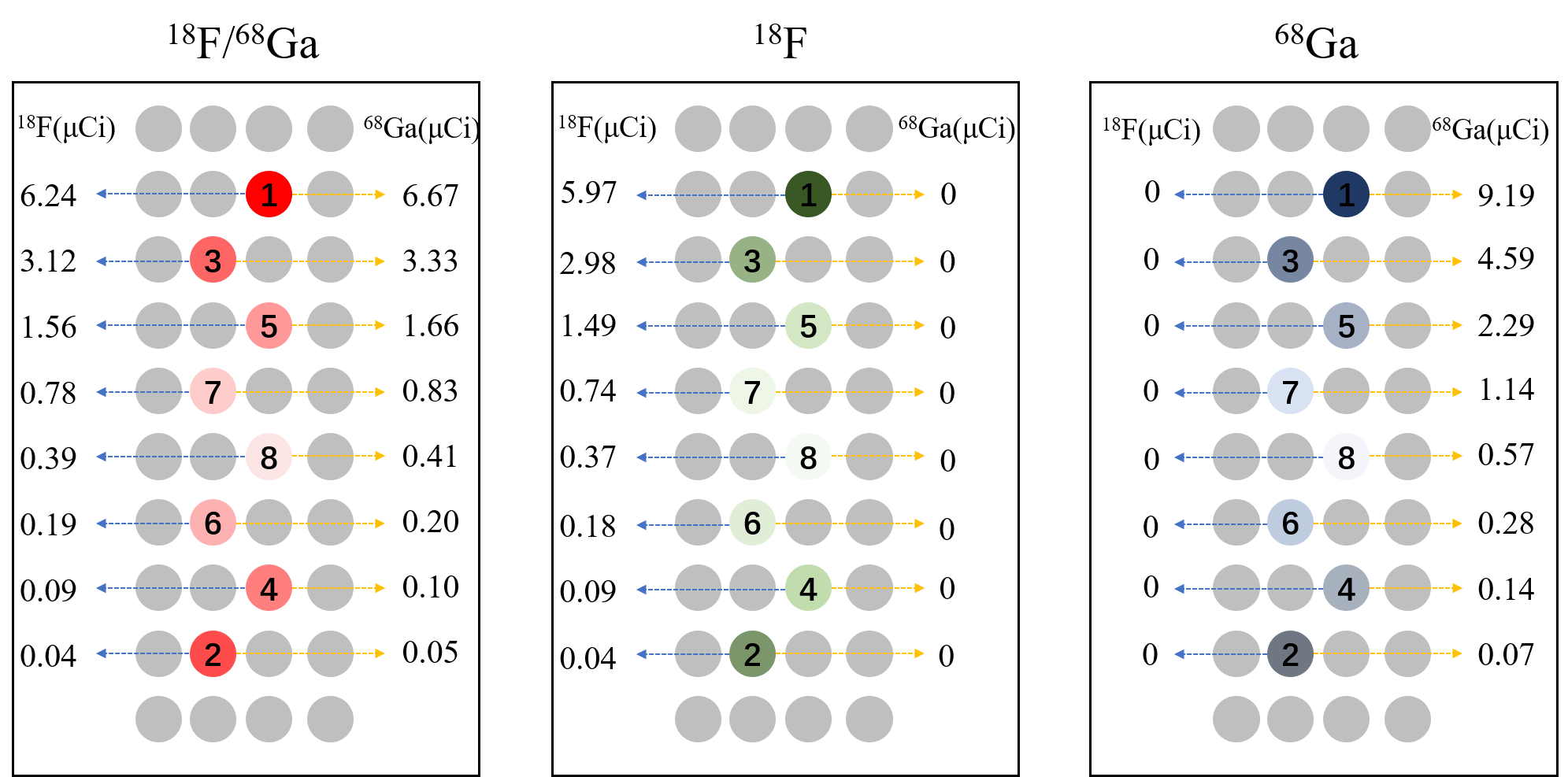}
	\caption{$^{18}$F and $^{68}$Ga distributions on each 96-well plate in the low-activity quantitation assessment.}
	\label{Fig2}
\end{figure}

As shown in Fig. \ref{Fig2}, the quantitation capability of the PNR-EM in $^{68}$Ga/$^{18}$F dual-isotope imaging was investigated using low-activity and high-activity 96-well plate phantoms, with each well measuring 7.0 mm in diameter and 11.0 mm in height. In the low-activity group, 8 wells were filled with a mixture of $^{68}$Ga and $^{18}$F solutions, with the highest activities of 6.24 $\mu$Ci for $^{18}$F and 6.67 $\mu$Ci for $^{68}$Ga. The activity was halved in each subsequent well, reaching 0.04 $\mu$Ci for $^{18}$F and 0.05 $\mu$Ci for $^{68}$Ga in the last well. The mixture ratio of the two solutions approached 1:1. Two additional 96-well plate phantoms filled with either single $^{18}$F or single $^{68}$Ga solutions at similar activity levels were used as control groups. Additionally, higher activities were injected into phantoms with the same layout to simulate lesion uptake. In the high-activity experimental group, which is not shown in Fig. \ref{Fig2}, the highest activity in the mixture was 44.58 $\mu$Ci for $^{18}$F and 45.10 $\mu$Ci for $^{68}$Ga, while the lowest activity was 0.34 $\mu$Ci for $^{18}$F and 0.35 $\mu$Ci for $^{68}$Ga. The data were acquired for 30 minutes and divided into three continuous 10-minute frames for analysis, considering the decays of nuclides. The single isotope distributions in control groups were reconstructed using classical MLEM on double coincidence as reference standards, while the $^{68}$Ga/$^{18}$F dual-isotope images were separated using PNR-EM. In addition, PNR-EM was also applied to the single-isotope scans for comparison. Consistency between the fitted quantitative curves was assessed using the maximum relative deviation (MRD) and average relative deviation (ARD). For $^{68}$Ga, MRD and ARD were evaluated over well activities of 0.5--50 $\mu$Ci. The MRD was additionally evaluated
within the low-activity subrange of 0.5--3 $\mu$Ci.

\subsubsection{Influence of PNR-EM on Image Resolution}
The potential degradation in image quality caused by the PNR-EM-based dual-isotope imaging strategy was assessed using Derenzo phantoms with hot-rod diameters of 2.4, 2.0, 1.6, 1.2, 1.0 and 0.7 mm. A total activity of 228 $\mu$Ci of $^{18}$F-FDG solution was injected into one Derenzo phantom, and 300 $\mu$Ci of $^{68}$Ga solution was injected into another phantom with the same structure. These two phantoms were placed at the center of the field of view (FOV) in turn and scanned for 15 minutes each. The single photon raw data from the two single-isotope scans were then mixed to generate a controlled simulated dual-isotope dataset. In this dataset, the external contours of the two Derenzo phantoms were aligned physically, whereas the internal radioactive rod patterns randomly overlapped. After coincidence detection, the imaging resolution of single isotopes reconstructed by MLEM was compared with that of the separated dual isotopes reconstructed by PNR-EM.

\subsubsection{Comparison of Dual-isotope Separation Methods}
\begin{figure}[!tbp]
\centering
	\includegraphics
	[width=0.6\hsize]
	{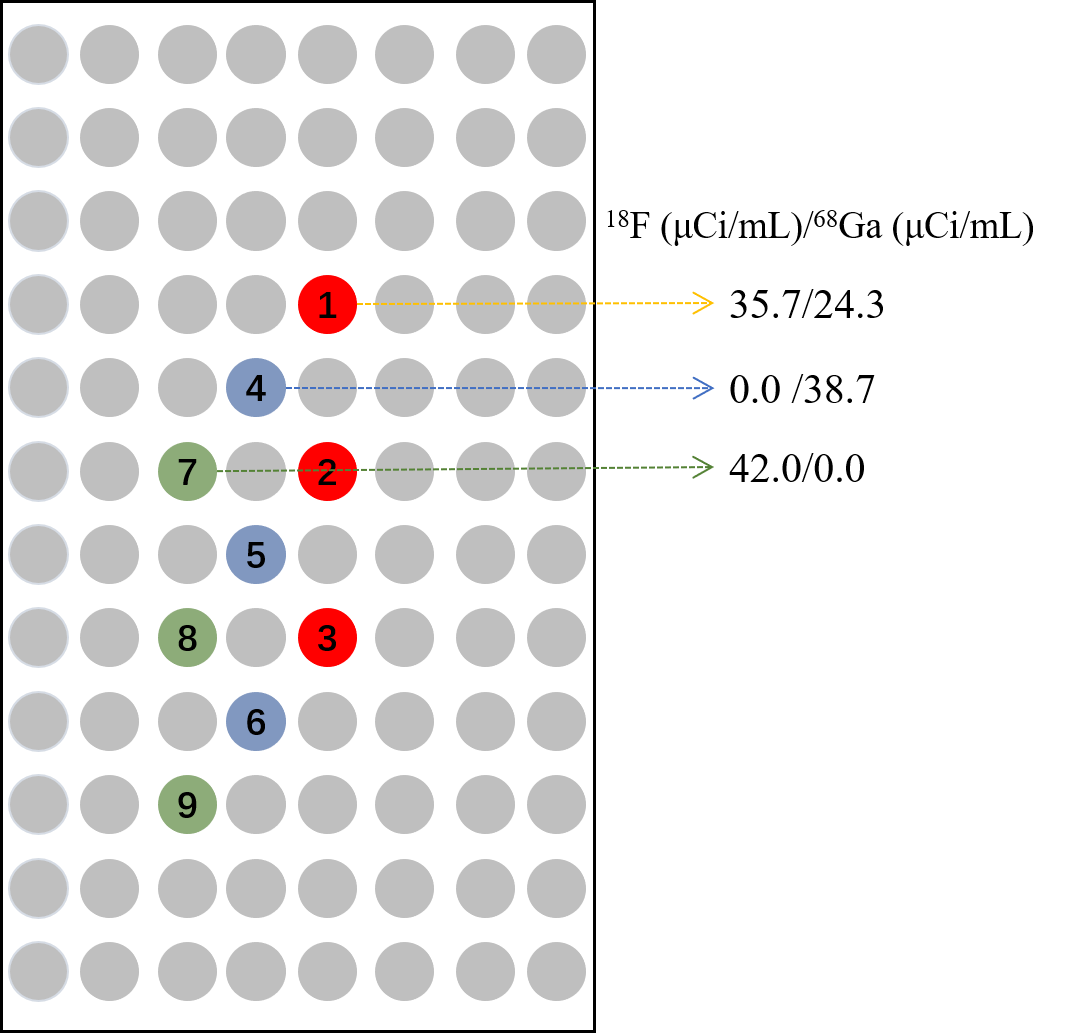}
	\caption{$^{18}$F and $^{68}$Ga distributions in the phantom experiment for algorithm performance comparison.}
	\label{Fig3}
\end{figure}

To validate the effectiveness of each component of the proposed PNR-EM, a new phantom was constructed using 9 wells of a 96-well plate. As shown in Fig. \ref{Fig3}, wells 1--3 contained a mixture of $^{18}$F and $^{68}$Ga, with concentrations of 35.7 $\mu$Ci/mL for $^{18}$F and 24.3 $\mu$Ci/mL for $^{68}$Ga at the beginning of the scan. Wells 4--6 contained only $^{68}$Ga at a concentration of 38.7 $\mu$Ci/mL, whereas wells 7--9 contained only $^{18}$F at a concentration of 42.0 $\mu$Ci/mL. The scan duration was 900 s. Except for classical MLEM and the proposed PNR-EM algorithms, the other compared algorithms are abbreviated in the form of $S$-$O$-$R$. Here $S$ denotes the separation strategy, including image-domain separation (Img) and domain-transform iterative separation (Iter). $O$ denotes the emitters considered in the reconstruction or separation objective function, where pure, non-pure and random components are denoted by P, N, R respectively. $R$ indicates whether the delayed window (DW) or consistent delay (CD) was used for random-triple correction. The compared methods are listed in Table \ref{tab1}. Additionally, the effect of TV regularization was evaluated using Iter-PN-DW and PNR-EM in this experiment. The number of iterations was fixed at 100 for the comparative experiments.

\begin{table}[!tbp]
\caption{Notation of abbreviations}
\label{tab1}
\centering
\begin{tabular}{lc}
\toprule
Abbreviations & Meanings \\
\midrule
MLEM &
\parbox[t]{5.7cm}{MLEM reconstruction using only double coincidence} \\

Img-N-DW \cite{RN22} &
\parbox[t]{5.7cm}{Image-domain dual-isotope separation after non-pure emitter reconstruction using delayed-window pre-corrected triples} \\

Img-N-CD &
\parbox[t]{5.7cm}{Image-domain dual-isotope separation after non-pure emitter reconstruction using consistent-delay pre-corrected triples} \\

Img-NR-DW &
\parbox[t]{5.7cm}{Image-domain dual-isotope separation after NR-EM reconstruction using delayed-window-extracted random triples} \\

Img-NR-CD &
\parbox[t]{5.7cm}{Image-domain dual-isotope separation after NR-EM reconstruction using consistent-delay-extracted random triples} \\

Iter-PN-DW \cite{RN29} &
\parbox[t]{5.7cm}{Pure--non-pure iterative dual-isotope separation using delayed-window pre-corrected triples} \\

PNR-EM &
\parbox[t]{5.7cm}{PNR-EM with Img-NR-CD initialization} \\

PNR-EM (R1) &
\parbox[t]{5.7cm}{PNR-EM with the random component initialized to ones} \\

PNR-EM (P1) &
\parbox[t]{5.7cm}{PNR-EM with the pure emitter initialized to ones} \\

PNR-EM (N1) &
\parbox[t]{5.7cm}{PNR-EM with the non-pure emitter initialized to ones} \\
\bottomrule
\end{tabular}
\end{table}

\subsubsection{In-vivo Evaluation of Tumor-Bearing Mice}
PNR-EM was further applied to in-vivo mouse data to evaluate its preclinical applicability. Three neuroendocrine tumor (NET)-bearing mice were injected with 159 $\mu$Ci of $^{18}$F-FDG, 146 $\mu$Ci of $^{68}$Ga-DOTA-NOC, and a mixture of $^{18}$F-FDG (163 $\mu$Ci) and $^{68}$Ga-DOTA-NOC (173 $\mu$Ci), respectively. Scan data were acquired for 15 minutes using the custom-built small-animal PET at 0.5-hour post tracer injection. The single-isotope PET images were reconstructed using MLEM and served as qualitative references for tracer uptake patterns. For the dual-isotope scan, the mixed or separated images were obtained using MLEM, Img-N-CD, Img-NR-CD, Iter-PN-DW, and PNR-EM.

\section{Results}
\subsection{Quantitation Assessment}
In the same PET imaging system, the relationship between reconstructed image values and radioactive activities, referred to here as quantitative curves, is expected to remain consistent after successful isotope separation. Therefore, for the $^{68}$Ga/$^{18}$F mixture with an activity ratio of approximately 1:1, the quantitative curves obtained by PNR-EM should be close to those obtained from the corresponding single-isotope scans. In addition, when prior knowledge establishes that only one isotope is present, namely an activity ratio of 1:0 or 0:1, PNR-EM is expected to approximately degenerate into conventional single-isotope MLEM reconstruction. Consequently, the quantitative curves obtained from dual-isotope separation by PNR-EM should theoretically be consistent with those obtained from single-isotope MLEM reconstruction and single-isotope PNR-EM separation.

\begin{figure*}[!htb]
\centering
	\includegraphics
	[width=0.9\hsize]
	{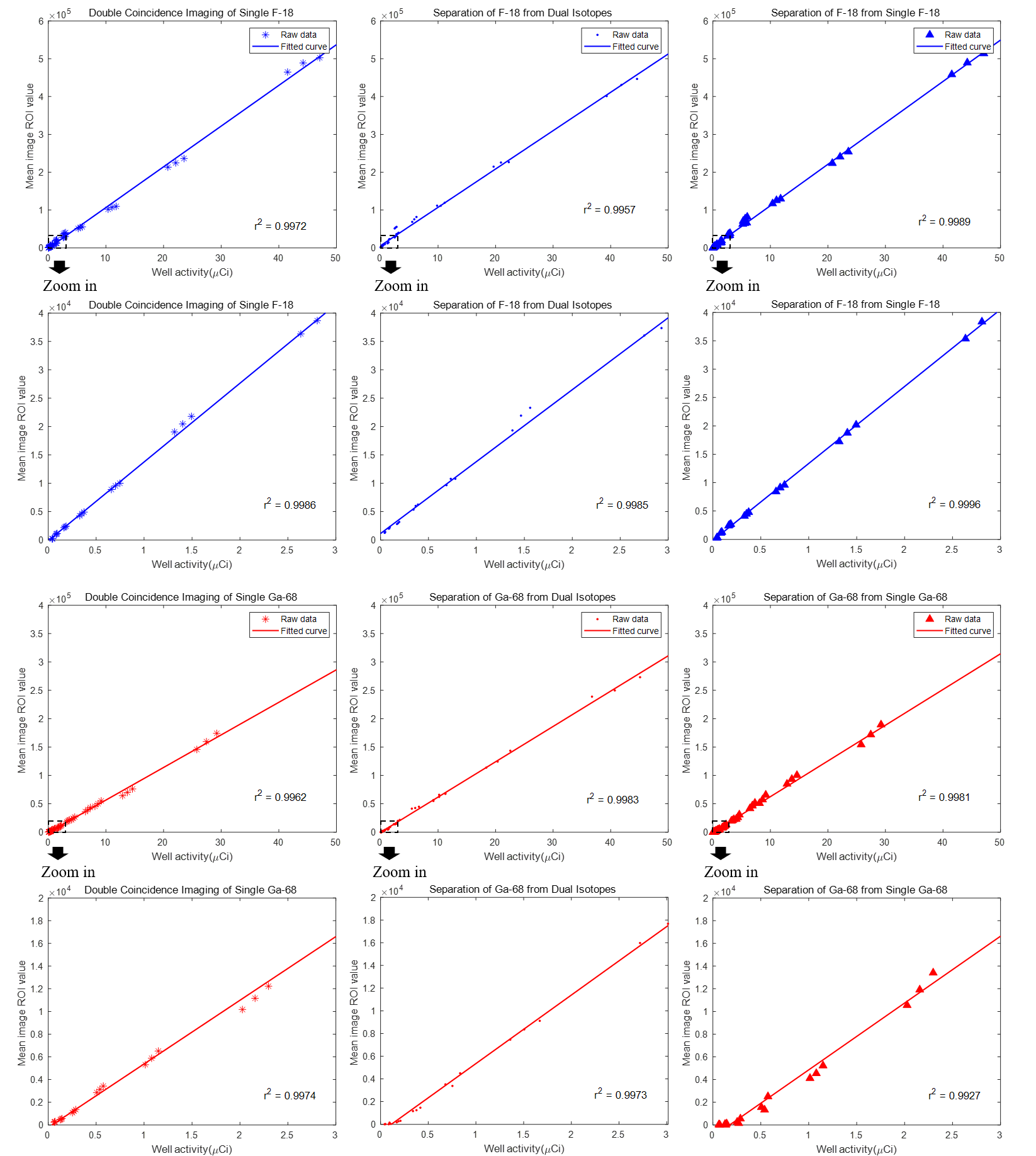}
	\caption{Relationship between reconstructed image values and well activities obtained by three methods. Left: MLEM reconstructions from single-tracer scans; Middle: PNR-EM separation from dual-tracer scans with a mixture ratio of ~1:1; Right: PNR-EM separation from single-tracer scans.}
	\label{Fig4}
\end{figure*}

Fig. \ref{Fig4} compares the quantitative curves obtained from MLEM reconstruction of single-isotope scans, PNR-EM separation of dual-isotope scans, and PNR-EM separation of single-isotope scans from left to right. Data from both low-activity and high-activity experimental groups are drawn in the same figure. As shown in the top row of Fig. \ref{Fig4}, the relationship between the reconstructed $^{18}$F’s voxel values and well activities presented a strong linearity under all three conditions, with $r^2\geq0.9957$. 
The quantitative curves obtained from $^{18}$F single-isotope MLEM reconstruction, $^{18}$F separation from dual-isotope scans, and $^{18}$F separation from single-isotope scans were highly consistent, with MRD of 9.81\% among the three curves. ARDs among the three conditions were all below 9.09\%. The second row of Fig. \ref{Fig4} further shows an enlarged view
of the quantitative curves for well activities below
3 $\mu$Ci, which is closer to the activity level of normal tissue. The fitted lines were retained from the full-activity-range analysis, whereas the coefficients of determination were recalculated using only the data points within this low-activity subrange without refitting. In this low-activity range, the relative deviation remained below 6.60\%. 

As the other part of dual-isotope imaging, the quantitative curves of $^{68}$Ga are shown in the bottom half of Fig. \ref{Fig4}. The linear fits yielded coefficients of determination greater than 0.99, indicating that PNR-EM achieved reliable quantitation for $^{68}$Ga/$^{18}$F dual-isotope imaging. Over the evaluated well-activity range of 0.5--50 $\mu$Ci, the MRD and ARD among the fitted quantitative curves of $^{68}$Ga were 8.86\% and 5.83\%, respectively. The biases of separated $^{68}$Ga images were generally lower than those of separated $^{18}$F images, because the reconstruction of the non-pure emitter in PNR-EM is jointly constrained by both prompt double coincidences and prompt triple coincidences, as shown in Eq. \eqref{eq8}. This complementary information from both data types improves the accuracy and stability of $^{68}$Ga quantification.
One notable difference between the curves for $^{18}$F and $^{68}$Ga is the vanishing voxel value in extremely low activity of $^{68}$Ga. As shown in the bottom of Fig. \ref{Fig4}, the voxel value of $^{68}$Ga approaches zero at an activity of ~0.15 $\mu$Ci (0.36 $\mu$Ci/mL), primarily due to the limited statistical count of triples. Within the low-activity evaluation subrange of 0.5--3 $\mu$Ci, the MRD among the fitted quantitative curves was also 8.86\%. The near-zero reconstructed value at approximately 0.15 $\mu$Ci lies below the lower activity limit used for the reported MRD and ARD calculations.

Overall, these results demonstrate that simultaneous $^{68}$Ga/$^{18}$F dual-tracer imaging with PNR-EM can provide quantitative accuracy comparable to conventional single-isotope PET within the evaluated activity range.

\subsection{Image Resolution}
\begin{figure}[!htb]
	\includegraphics
	[width=1.0\hsize]
	{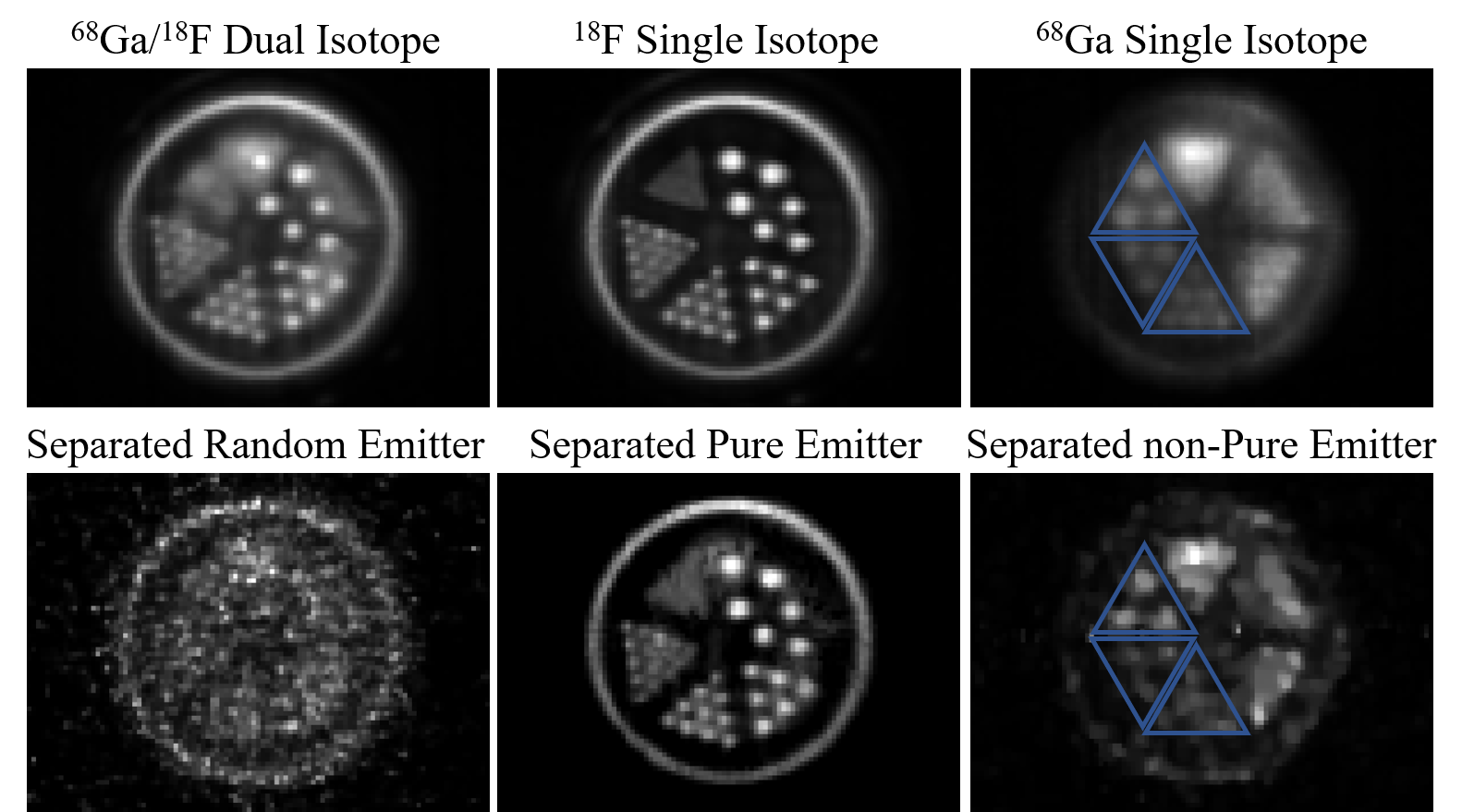}
	\caption{Comparison of image resolution among mixed dual-isotope imaging by MLEM, single-isotope imaging by MLEM, and dual-isotope separation by PNR-EM using Derenzo phantoms.}
	\label{Fig5}
\end{figure}

This study conducted a $^{68}$Ga/$^{18}$F dual-isotope imaging experiment on Derenzo phantoms to evaluate the impact of the proposed algorithm on image resolution. The top row of Fig. \ref{Fig5} presents the mixed dual-isotope images reconstructed by MLEM, the $^{18}$F single-isotope reference images and the $^{68}$Ga single-isotope reference images in sequence. In the mixed dual-isotope MLEM image, the $^{68}$Ga signal was largely overwhelmed by the stronger $^{18}$F signal, making the $^{68}$Ga rod pattern difficult to distinguish in the mixed image and reducing its apparent separability.
The second row of Fig. \ref{Fig5} displays the random component, pure-emitter component ($^{18}$F), and non-pure-emitter component ($^{68}$Ga) separated by PNR-EM sequentially. Compared with the corresponding single-isotope references, the separated $^{18}$F image retained a visual resolution between 1.0 and 1.2 mm, while the separated $^{68}$Ga image maintained an approximate visual resolution of 2.0 mm. The visual analysis suggests that PNR-EM introduces only limited degradation in image resolution for simultaneous $^{68}$Ga/$^{18}$F dual-isotope imaging and separation.

\begin{figure}[!htb]
	\includegraphics
	[width=1.0\hsize]
	{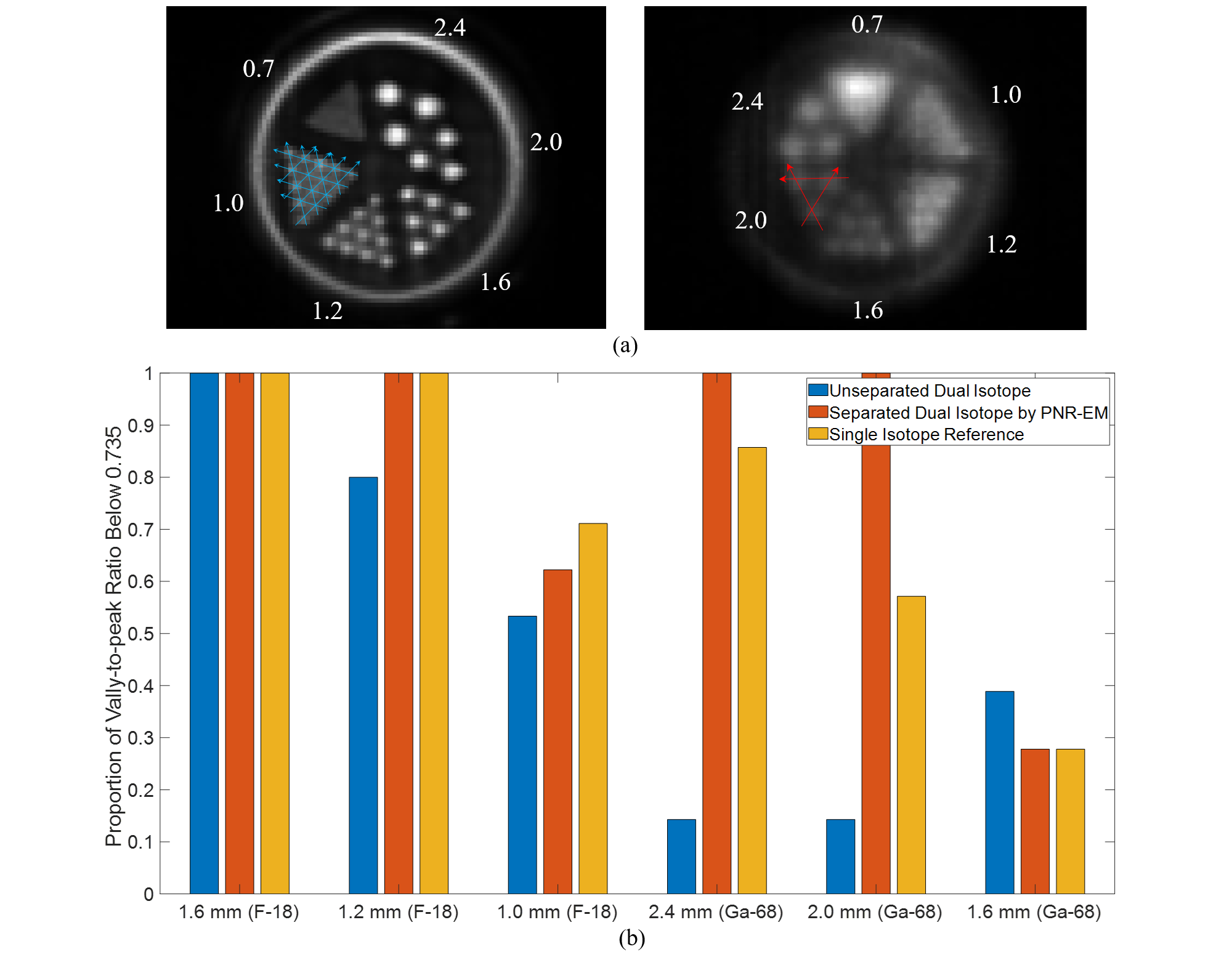}
	\caption{Quantitative evaluation of the effects of reconstruction/separation algorithms on image resolution.}
	\label{Fig6}
\end{figure}

Subsequently, a quantitative analysis was conducted according to the paper \cite{RN40}. As shown in Fig. \ref{Fig6}(a), the valley-to-peak ratio (VPR) between each background valley and its adjacent hot rod was calculated along the straight-line direction. A VPR below 0.735 was considered as the criterion for source distinguishability according to the Rayleigh criterion, and the proportion of VPR values satisfying this criterion was calculated. Fig. \ref{Fig6}(b) summarizes the proportion of VPR in the regions close to the resolution limits, including the 1.6, 1.2, and 1.0 mm regions for $^{18}$F and the 2.4, 2.0, and 1.6 mm regions for $^{68}$Ga Derenzo image.

As illustrated in Fig. \ref{Fig6}(b), for the MLEM-reconstructed dual-isotope mixture, the VPR proportion in the 1.2 mm region of the $^{18}$F image was only 0.80, which was lower than the 0.90 threshold used in reference \cite{RN40}. This indicates that the mixed double-coincidence-based MLEM image did not reach a resolution of 1.2 mm. In contrast, PNR-EM achieved a VPR proportion of 1.00 in the 1.2 mm region, consistent with the $^{18}$F single-isotope reference. This result suggests that the separated $^{18}$F image obtained by PNR-EM preserved a resolution of approximately 1.2 mm. The difference between PNR-EM and the single-isotope reference became more evident in the 1.0 mm region, where the VPR proportion was 0.62 for PNR-EM and 0.71 for the single-isotope reference. PNR-EM exhibits this decrease in the proportion of VPR due to the existence of two additional sets of peaks and valleys that do not satisfy the Rayleigh criterion compared to the reference image. However, since both references and PNR-EM separations yielded a VPR proportion below 0.9 threshold in the 1.0 mm region, neither image was considered to reach a resolution of 1.0 mm. Therefore, despite minor degradation in fine structures, the practical resolution level of the separated $^{18}$F image remained consistent with that of the single-isotope reference.

For $^{68}$Ga images, PNR-EM achieved VPR passing proportions of 1.00 in both the 2.4 and 2.0 mm regions, indicating that 2.0-mm rods were resolved under the adopted VPR criterion, consistent with the visual observation in Fig. \ref{Fig5}. The corresponding single-isotope reference yielded passing proportions of approximately 0.86 and 0.57 in the 2.4 and 2.0 mm regions, respectively, both below the 0.90 threshold. Thus, PNR-EM did not show reduced VPR-based rod separability relative to the single-isotope reference in these regions.
This observation should be interpreted cautiously. Both the 2.0 and 2.4 mm regions of the Derenzo phantom contain only a limited number of rods, which may introduce statistical uncertainty into the VPR proportion. Furthermore, differences in reconstruction-induced contrast and noise may affect the threshold-based VPR assessment \cite{RN40}. These factors could contribute to the observed differences without necessarily indicating a reduction in positron-range blurring.
In the 1.6 mm region, the MLEM-reconstructed mixed image showed the highest VPR proportion. This was likely caused by the spatial overlap between the 1.2 mm region of the $^{18}$F phantom and the 1.6 mm region of the $^{68}$Ga phantom, which artificially enhanced the apparent $^{68}$Ga structure in the unseparated mixed image.

\subsection{Comparison between Different Separation Methods}
\begin{figure}
	\includegraphics
	[width=1.0\hsize]
	{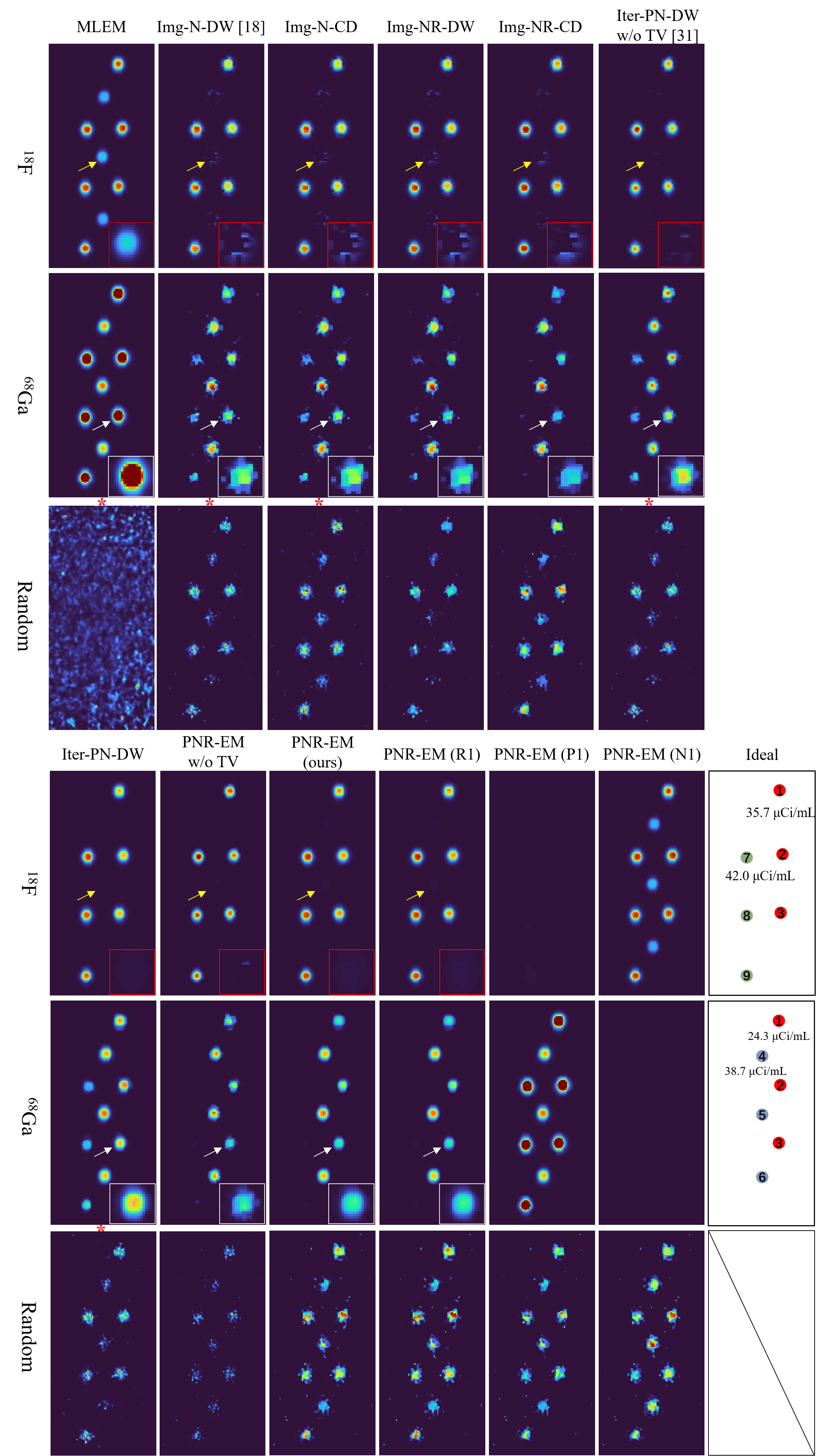}
	\caption{Qualitative comparisons of $^{68}$Ga/$^{18}$F dual-isotope separation using PNR-EM and other methods. Asterisk \textcolor{red}{*} indicates that random triples are not reconstructed jointly with isotope separation, but are instead estimated from delayed events using MLEM individually.}
	\label{Fig7}
\end{figure}

Fig. \ref{Fig7} compares the performance of different methods for simultaneous $^{18}$F/$^{68}$Ga dual-isotope separation, highlighting the advantages of PNR-EM in both random-triple correction and radionuclide separation.

Regarding random-triple correction, the $^{68}$Ga images obtained using Img-NR-CD were compared with those obtained from Img-N-CD, and the $^{68}$Ga images obtained using PNR-EM were compared with those obtained from Iter-PN-DW as well. The visual results show that modeling the random-triple component within the iterative objective function provides clear advantages over conventional projection-domain pre-correction based on direct subtraction, especially under low-count triple-coincidence conditions. After NR-EM correction, residual pure-emitter signals in the separated non-pure-emitter images were visibly suppressed.

The comparison between Img-NR-DW and Img-NR-CD further shows that the use of consistent delay effectively reduced the residual $^{18}$F signal in the separated $^{68}$Ga images, contributing to more accurate image quantification through effective CRT correction. In contrast, Img-N-CD showed no obvious improvement over Img-N-DW, mainly because projection-domain subtraction is unstable when the triple-coincidence statistics are low \cite{RN31}. These observations support the necessity of incorporating the random component into the objective function and performing random-triple correction within the iterative separation process. By combining NR-EM with consistent delay, PNR-EM produced separated isotope images that were visually closer to the known isotope distributions, as shown in Fig. \ref{Fig7}.

In terms of dual-isotope separation, affected by the low count of triples, image-domain methods such as Img-NR-CD produced $^{68}$Ga distributions that differed substantially from the known distributions in both image contour and activity distribution. This inaccuracy further caused overestimation or underestimation in subsequent pure emitter separation. Nonphysical negative values occurred in the $^{18}$F images obtained by image-domain subtraction. The yellow arrows and red-bordered insets in Fig. \ref{Fig7} highlight the relevant regions for comparison across methods. A similar phenomenon was also observed in the in-vivo results shown in Fig. \ref{Fig8}. According to Eq. \eqref{eq8}, by jointly modeling the pure, non-pure, and random components, PNR-EM enables the non-pure emitters $f_B$ to be optimized using statistically reliable double-coincidence data together with triple-coincidence data, rather than relying solely on triple-coincidence data as in image-domain separation. During the iterative process, optimizing the distribution of the non-pure $^{68}$Ga isotope also leads to a more accurate distribution of pure $^{18}$F and random components in turn. Therefore, Fig. \ref{Fig7} shows that dual-isotope images separated by PNR-EM and Iter-PN-DW have image contours closer to the known isotope distributions. In addition, TV regularization further suppressed noise and contour degradation caused by low triple-coincidence statistics on the basis of the above iterative separation. Compared with Iter-PN-DW w/o TV or PNR-EM w/o TV, the TV-regularized algorithms tended to produce smoother $^{68}$Ga images, as indicated by the white arrows and the zoomed regions of corresponding $^{68}$Ga subgraphs in Fig. \ref{Fig7}.

As mentioned earlier, in order to ensure and accelerate the convergence of the proposed algorithm, the separation results of 20 iterations using Img-NR-CD are adopted as initial values for PNR-EM. Therefore, the influence of different initial values on separation performance was also investigated. Comparing PNR-EM with PNR-EM (R1) shows that initializing the random component with an all-one image had little influence on the final separation results. However, when either $^{18}$F or $^{68}$Ga image was initialized as an all-one matrix while the other isotope image to be separated was initialized using Img-NR-CD result after 20 iterations, the final separation became incorrect. One possible reason is the large difference in image-value scales among the components in this phantom experiment. The voxel values of the isotope images were on the order of $10^5$ which is much larger than 1, whereas those of the random component were on the order of $10^1$--$10^2$, much smaller than those of the isotope images. Therefore, initializing only the random component with ones had a relatively limited influence on convergence. Conversely, when one of the isotopes (e.g., $f_B$) was initialized as an all-one matrix instead, then $f_A^k(j)\gg(1-\beta(j))f_B^k(j)$ during the early iterations. In this case, the update of $f_A$ approximately degenerates into a TV-regularized double-coincidence reconstruction update, as shown in Eq. \eqref{eq9}. As a result, the separated $^{18}$F image of PNR-EM (N1) in Fig. \ref{Fig7} resembled the mixed double-coincidence image, while the separated $^{68}$Ga image approached zero.
\begin{equation}\label{eq9}
f_A^{k + 1}(j) \approx \frac{{f_A^k(j)}}{{\sum\limits_i {{a_{ij}}} }+\lambda U_A^k(j)}\sum\limits_i {\frac{{{a_{ij}}double_p(i)}}{{\sum\limits_{j'} {{a_{ij'}}f_A^k(j')} }}}
\end{equation}

Furthermore, since the isotope distribution in the 9-well phantom was known, reference images were available for quantitative evaluation. As shown in Table \ref{tab2}, the incorporation of NR-EM and Consistent Delay substantially improved the PSNR and SSIM values and reduced RMSE for $^{68}$Ga separation. However, due to limitations of image-domain separation, no substantial improvement was observed in the quality of $^{18}$F images obtained using Img-NR-CD. In contrast, iterative separation through Iter-PN-DW effectively reduced statistical fluctuations in $^{18}$F separation caused by insufficient triple-coincidence counts. Hence, Iter-PN-DW demonstrated excellent performance in separating $^{18}$F images as shown in Table \ref{tab2}. PNR-EM uses the complementary information provided by the double, triple, and delayed/random triple measurements to simultaneously address isotope separation and random-triple contamination within the same iterative framework. Consequently, the proposed algorithm, including PNR-EM and PNR-EM (R1), achieved the best overall performance. For $^{18}$F separation, compared with the best non-PNR baseline Iter-PN-DW, PNR-EM (R1) reduced the RMSE of the separated $^{18}$F image by 2.4\%. For $^{68}$Ga separation, compared with the best non-PNR baseline Img-NR-CD, PNR-EM increased the PSNR by 4.67 dB and reduced the RMSE by 41.4\%.
These results support the combined benefits of joint
statistical modeling, complementary utilization of prompt double, prompt triple, and delayed triple measurements, and TV regularization for dual-isotope separation.

\begin{table*}[!htb]
	\caption{Quantitative comparisons of various PET separation methods using a 9-well phantom}
	\label{tab2}
    \centering
	\begin{tabular*}{16cm} {@{\extracolsep{\fill} } lcccccc}
		\toprule
		\multirow{2}{2cm}{Methods} & \multicolumn{3}{c}{$^{18}$F} & \multicolumn{3}{c}{$^{68}$Ga} \\
		\cmidrule(lr){2-4}\cmidrule(lr){5-7}
		& PSNR (dB, $\uparrow$) & SSIM($\uparrow$) & RMSE($\downarrow$) & PSNR (dB, $\uparrow$) & SSIM($\uparrow$) & RMSE($\downarrow$)\\
		\midrule
		MLEM  & 29.13 & 0.939 & 0.0349 & 20.69 & 0.893 & 0.0923\\
		Img-N-DW & 33.15 & 0.962 & 0.0220 & 28.97 & 0.922 & 0.0356 \\
		Img-N-CD & 33.35 & 0.962 & 0.0215 & 29.05 &	0.923 &	0.0353 \\
		Img-NR-DW &	33.41 &	0.964 &	0.0213 & 29.65 & 0.931 & 0.0329 \\
		Img-NR-CD &	32.86 &	0.965 &	0.0227 & 30.19 & 0.952 & 0.0309\\
		Iter-PN-DW w/o TV &	30.39 &	0.972 &	0.0302 & 27.00 & 0.921 & 0.0447\\
		Iter-PN-DW & 33.49 & \underline{\textit{0.974}} & 0.0211 & 27.92 & 0.931 & 0.0402\\
		PNR-EM w/o TV & 30.81 & 0.972 &	0.0288 & 30.64 & $\mathbf{0.955}$ & 0.0294\\
		PNR-EM & \underline{\textit{33.69}} & \underline{\textit{0.974}} & \underline{\textit{0.0207}} & $\mathbf{34.86}$ & $\mathbf{0.955}$ & $\mathbf{0.0181}$\\
		PNR-EM (R1) & $\mathbf{33.72}$ & $\mathbf{0.975}$ & $\mathbf{0.0206}$ &\underline{\textit{34.78}} & \underline{\textit{0.953}} & \underline{\textit{0.0182}}\\
		PNR-EM (P1) & 19.17 & 0.056 & 0.1101 & 20.54 & 0.894 & 0.0940\\
		PNR-EM (N1) & 28.85 & 0.939 & 0.0361 & 7.37 & 0.021 & 0.4281\\
		
		\bottomrule
	\end{tabular*}
    \vspace{2pt}
    \parbox{\linewidth}{
        \footnotesize
        \raggedright
        \textit{Note:} $\mathbf{Bold}$ and \underline{\textit{italic underlined}} fonts indicate the optimal and suboptimal results, respectively.}
\end{table*}

\subsection{Tumor-bearing Mice Imaging}
\begin{figure}[!htb]
	\includegraphics
	[width=1.0\hsize]
	{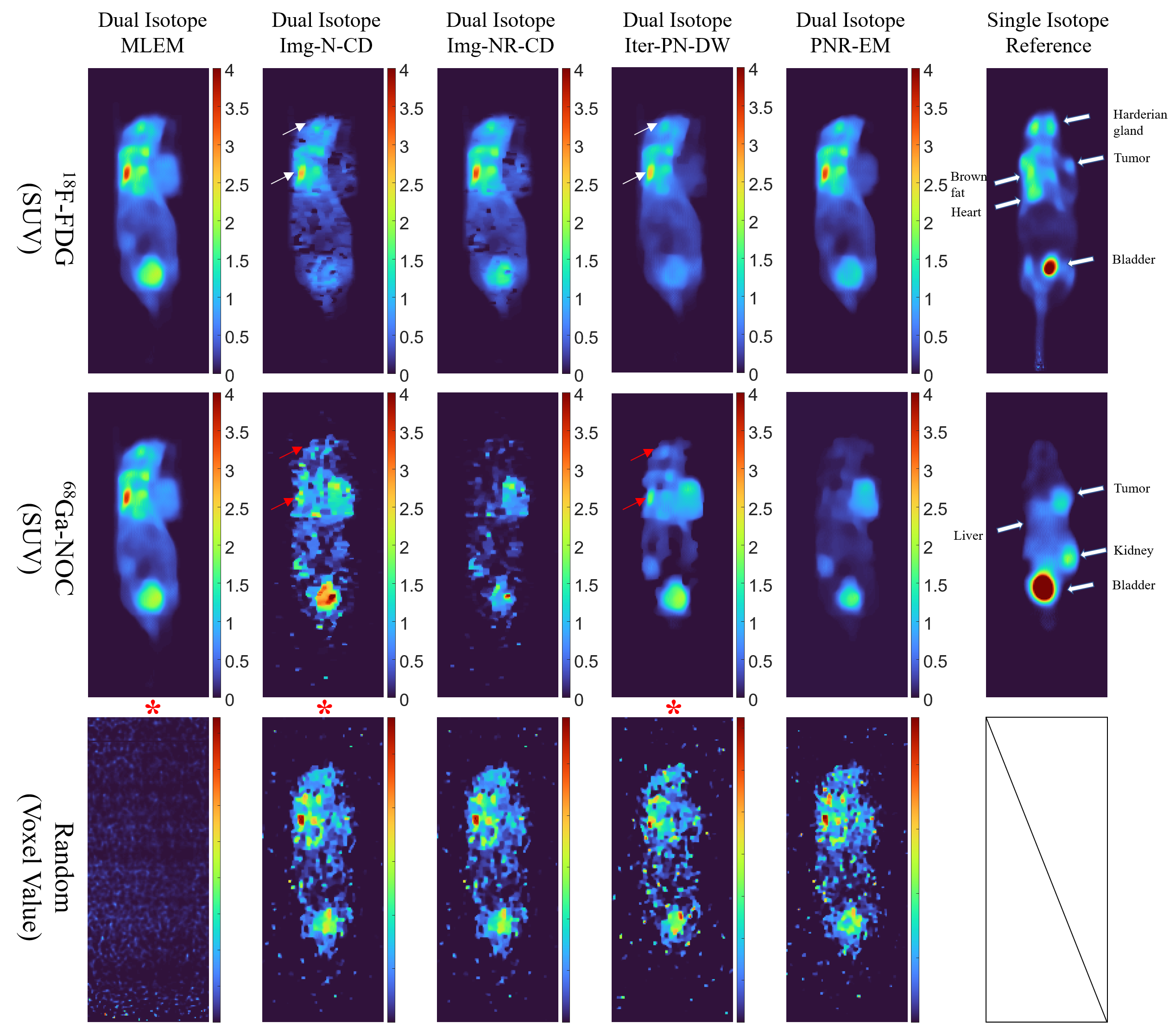}
	\caption{Separated $^{18}$F/$^{68}$Ga dual-isotope mouse PET images using different methods. Asterisk \textcolor{red}{*} indicates that random triples are not reconstructed jointly with isotope separation, but are instead estimated from delayed events using MLEM individually.}
	\label{Fig8}
\end{figure}

PNR-EM was further evaluated using in-vivo NET-bearing mouse data and compared with conventional double-coincidence-based MLEM and three representative separation methods. In addition, single-isotope imaging of $^{18}$F-FDG and $^{68}$Ga-DOTA-NOC was performed on two other NET-bearing mice respectively and used as qualitative reference images. As shown in Fig. \ref{Fig8}, in the single $^{18}$F-FDG imaging, relatively high uptake was observed in the bladder, heart, brown fat, Harderian gland, and the subcutaneous tumor on the right side, with an $SUV_{\max}$ of 0.7 in the tumor region. In contrast, in the single $^{68}$Ga-DOTA-NOC reference image, apart from the bladder, high uptake was mainly observed in the tumor and kidneys, with an $SUV_{\max}$ of 1.8 in the tumor region. When the dual tracers were simultaneously injected into the same mouse and reconstructed using MLEM algorithm, the resulting image represented a superposition of both tracers and could not differentiate their respective spatial distributions.

By utilizing triple coincidences as unique information of non-pure emitters ($^{68}$Ga in this study), dual-isotope separation can theoretically be achieved from mixed images. However, even after projection-domain pre-correction of random triples, such as Img-N-CD, elevated false-positive uptake were observed in Harderian gland, heart, and brown fat regions of the separated $^{68}$Ga-DOTA-NOC image, as indicated by red arrows. In these regions, the $SUV_{\mathrm{mean}}$ was 1.4 and the $SUV_{\max}$ reached 2.4. This contamination also led to underestimation of $^{18}$F-FDG concentration, as indicated by white arrows. A similar phenomenon was observed in the dual-isotope iterative separation via Iter-PN-DW.

By incorporating the random-triple component together with the non-pure emitter in the objective function, NR-EM effectively reduced random-triple contamination. The $SUV_{\mathrm{mean}}$ in the heart and brown fat regions decreased to approximately 0.1 in the separated $^{68}$Ga-DOTA-NOC image. In addition, the estimated random component showed a spatial distribution similar to that of the double-coincidence image, suggesting that CRTs were the dominant random triple in $^{68}$Ga/$^{18}$F dual-isotope imaging, consistent with previous findings \cite{RN34}. However, image-domain separation after NR-EM still suffered from statistical fluctuations in the pure-emitter image, which is a common limitation of image-domain subtraction methods \cite{RN24}.

Based on a joint Poisson distribution of pure, non-pure, and random emitters, PNR-EM simultaneously estimated all three components within an expectation-maximization framework. The tumor-region $SUV_{\max}$ values separated by PNR-EM were 0.5 for $^{18}$F-FDG and 1.2 for $^{68}$Ga-DOTA-NOC, respectively. The single-isotope reference images were acquired from different mice and therefore provided qualitative uptake-pattern comparisons rather than quantitative ground truth for the dual-tracer scan. The PNR-EM results showed spatial distributions more consistent with the corresponding reference uptake patterns than those obtained by Iter-PN-DW. Compared with Img-NR-CD, PNR-EM produced separated images with lower visible noise and closer quality to conventional PET imaging. These findings support the feasibility of simultaneous $^{68}$Ga-DOTA-NOC and $^{18}$F-FDG separation in vivo.

\section{Discussion}
$^{68}$Ga/$^{18}$F dual-isotope PET imaging has attracted increasing interest in neuroendocrine tumor research because it has the potential to provide complementary metabolic and receptor-specific information within a single scan.
Prompt-gamma-based imaging theoretically enables the separation of $^{68}$Ga and $^{18}$F by exploiting the difference in photons between non-pure and pure positron emitters.
However, the prompt-gamma branching ratio of $^{68}$Ga is extremely low, and the resulting triple-coincidence events are therefore highly limited in count statistics. More severely, these weak triple-coincidence signals are heavily contaminated by random triple coincidences, which substantially compromises their direct use for isotope separation. Consequently, in previous studies \cite{RN25,RN26,RN27}, $^{68}$Ga was often imaged as a pure positron emitter. In contrast, this study explicitly considers the impact of random triples in $^{68}$Ga/$^{18}$F dual-isotope imaging and proposes a random-triple estimation strategy based on consistent delay as the preparatory step for PNR-EM. Then, considering the low-count characteristics of triple coincidences, a joint Poisson model is further developed to incorporate pure and non-pure isotopes along with random components within a unified reconstruction framework. In this model, the random component is introduced as an auxiliary image-domain variable to account for random triple contamination, although it does not correspond to a physically existing radioactive distribution. The non-pure $^{68}$Ga component is jointly constrained by prompt double and prompt triple coincidences, and its noise level is further controlled with TV regularization. Through iterative EM updates, PNR-EM enables stable separation of $^{68}$Ga/$^{18}$F dual isotopes under low triple-coincidence statistics. Phantom experiments supported the quantitative performance of PNR-EM under the evaluated conditions, while tumor-bearing mouse imaging demonstrated the feasibility of simultaneous $^{68}$Ga-DOTA-NOC and $^{18}$F-FDG separation in vivo. To the best of our knowledge, this is the first study to demonstrate prompt-gamma-related triple coincidences for simultaneous imaging and separation of $^{68}$Ga/$^{18}$F dual isotopes.

In terms of experimental design, this study used four complementary experiments to evaluate the proposed method. A 96-well plate phantom was designed for quantitative imaging evaluation, a dual-tracer Derenzo phantom was used to assess the influence of PNR-EM on image resolution, a 9-well phantom containing both mixed and single-isotope regions was constructed to compare different separation methods, and tumor-bearing mouse experiments were performed to evaluate the in-vivo feasibility of the proposed method. The animal evaluation was exploratory, with one mouse per tracer condition and no matched single-isotope ground truth for the dual-tracer mouse. Larger animal studies with independent tracer-specific reference measurements are needed to assess in vivo quantitative accuracy.
In these experiments, an approximately 1:1 activity ratio was used as a controlled setting for the initial evaluation of PNR-EM. However, the administered activities of $^{18}$F-FDG and $^{68}$Ga-DOTA-NOC may vary across clinical protocols \cite{RN42,RN43}, and the relative tracer activities within the imaging field of view also depend on their biodistributions. Therefore, further evaluation under different activity ratios is needed to assess the robustness and broader applicability of the proposed method.

One limitation of the current model is the conditional-independence approximation used to factorize the complete-data likelihood. Given the image parameters $f_A$, $f_B$, and $f_r$, the corresponding count contributions are treated as independent Poisson variables to simplify the likelihood formulation and EM updates. In reality, random-triple formation depends on the underlying isotope distributions, and CRTs involve a true annihilation pair from one decay and an additional photon from another decay. The current model estimates $f_r$ as an auxiliary component without explicitly parameterizing its dependence on $f_A$ and $f_B$ through the random-triple formation process. Potential correlations among prompt and delayed coincidence counts arising from shared event data are not explicitly
modeled. Nevertheless, the phantom and in vivo results support the practical utility of the proposed reconstruction under the evaluated conditions. More accurate statistical models that explicitly describe random-triple formation may further improve the physical fidelity of the reconstruction framework.

The choices of initial values and appropriate regularization are other crucial topics in the future work. This study demonstrates that unlike MLEM, the PNR-EM algorithm cannot be initialized arbitrarily, as inappropriate initial values may lead to incorrect isotope separation. Therefore, this paper provides a feasible approach for determining initial values by using the results of Img-NR-CD with fewer iterations. Although this initialization strategy produced stable separation results for $^{68}$Ga/$^{18}$F dual-isotope imaging, it also increased the overall algorithmic complexity, which may limit practical clinical application. Therefore, future research will focus on developing more convenient and robust initialization strategies.
In addition, with the aid of double-coincidence information, classical TV regularization achieved acceptable noise suppression in this work. However, TV regularization may introduce cartoon-like artifacts in the separated isotope images, especially in the low-count $^{68}$Ga images, because of its sparse-gradient assumption. Future work will explore adaptive or learning-based regularization strategies to further improve image quality by combining the advantages of model-driven reconstruction and data-driven priors.

\section{Conclusion}
$^{68}$Ga/$^{18}$F dual-isotope imaging shows promising potential for the diagnosis and management of neuroendocrine tumors. Although prompt-gamma-based imaging techniques are anticipated to facilitate simultaneous $^{68}$Ga/$^{18}$F dual-isotope imaging, they face challenges related to low count statistics and random contamination from triple coincidences. To address these challenges, this paper employs a consistent-delay strategy for random-triple estimation, and further proposes a novel joint Poisson model that incorporates pure emitter, non-pure emitter, and random-triple components. Within this framework, the proposed PNR-EM estimates the three components simultaneously through EM iteration, suppresses random-triple contamination, and improves the stability of separation by jointly using prompt double, prompt triple, and delayed/random triple measurements with TV regularization.
Qualitative and quantitative evaluations using physical phantoms and tumor-bearing mice demonstrate that PNR-EM can achieve simultaneous separated imaging of $^{68}$Ga and $^{18}$F while maintaining quantitative performance and image quality close to those of corresponding single-isotope acquisitions under the evaluated conditions. 
Moreover, the applicability of the proposed framework may extend beyond $^{68}$Ga to other non-pure emitters with low prompt-gamma emission probabilities and substantial random-triple contamination. This work provides a methodological basis for the broader development and application of prompt-gamma-enabled dual-isotope PET imaging.




\noindent\textbf{Ethical statement} All animal procedures were approved by the Institutional Animal Care and Use Committee of Air Force Medical University (Fourth Military Medical University) with approval No. 251623.


\bibliographystyle{IEEEtran}
\bibliography{reference_0909}

\end{document}